\documentclass[twocolumn,showpacs,amsmath,amssymb,prl,floatfix]{revtex4-1}

\usepackage{graphicx}
\usepackage{dcolumn}
\usepackage{bm}
\usepackage{bbm}
\usepackage{multirow,amssymb,amsbsy,amsmath}
\usepackage{stmaryrd}

\newcommand{\ddim}{\udelta\kern0.1em}

\newcommand{\beikonst}[2]{\left( #1 \right)_{\kern-0.2em #2}}
\newcommand{\tr}[2][]{\text{Tr}_{#1}\left\{#2\right\}}

\newcommand{\trtxt}[2][]{\text{Tr}_{#1}\{#2\}}

\begin{document}


%
%

\title{Variational principle for steady states of dissipative quantum
  many-body systems}

\author{Hendrik Weimer}%
\email{hweimer@itp.uni-hannover.de}
\affiliation{Institut f\"ur Theoretische Physik, Leibniz Universit\"at Hannover, Appelstr. 2, 30167 Hannover, Germany}

\date{\today}%

\begin{abstract}

  We present a novel generic framework to approximate the
  non-equilibrium steady states of dissipative quantum many-body
  systems. It is based on the variational minimization of a suitable
  norm of the quantum master equation describing the dynamics. We show
  how to apply this approach to different classes of variational
  quantum states and demonstrate its successful application to a
  dissipative extension of the Ising model, which is of importance to
  ongoing experiments on ultracold Rydberg atoms, as well as to a
  driven-dissipative variant of the Bose-Hubbard model. Finally, we
  identify several advantages of the variational approach over
  previously employed mean-field-like methods.

\end{abstract}


\maketitle

From Fermat's principle of geometric optics, to Hamilton's principle
of classical mechanics, to the maximum entropy principle of
statistical physics, variational formulations are instrumental in our
understanding of physical theories. Similarly, variational principles
hold great power for the derivation of approximative solutions to
computationally hard quantum many-body problems, as demonstrated by
the successes of density functional theory \cite{Kohn1999} and the
matrix product state formalism \cite{Schollwock2011}. Here, we
establish a variational principle for the steady states of dissipative
quantum systems and demonstrate its effectiveness to tackle otherwise
intractable problems.

Dissipative quantum many-body systems have recently received great
attention, in particular following the discovery that dissipation can
be a useful resource in the controlled preparation of quantum-many
body states
\cite{Diehl2008,Verstraete2009,Weimer2010,Diehl2010,Alharbi2010,Barreiro2011,Krauter2011,Watanabe2012,Rao2013,Carr2013a,Weimer2013a,Otterbach2014}. The
interplay between coherent and dissipative dynamics can lead to a rich
variety of novel physical phenomena, such as interaction-mediated
laser cooling \cite{Huber2012,Zhao2012}, dissipative binding
mechanisms \cite{Ates2012,Lemeshko2013a}, and phase transitions in the
non-equilibrium steady state of the dynamics
\cite{Kasprzak2006,Amo2009,Hartmann2010,Baumann2010,Nagy2010,Diehl2010a,Tomadin2011,Lee2011,Kessler2012,Honing2012,Honing2013,Horstmann2013,Torre2013,Sieberer2013,Qian2013,Lee2013,Carr2013,Joshi2013,Malossi2014,Marcuzzi2014,Lang2014}. However,
the theoretical understanding of these systems is still fairly
limited, especially because of the lack of a concept corresponding to
the partition function in equilibrium systems, from which all
quantities of interest can be derived. Additionally, dissipative
quantum many-body system exhibit the same computational complexity as
equilibrium systems, with the Hilbert space dimension growing
exponentially with the size of the system.

In this Letter, we establish a variational principle for steady states
of dissipative quantum many-body systems. Crucially, our approach is
completely generic and allows to analyze arbitrary dissipative models
using arbitrary variational quantum states. The key element is the
minimization of a suitable norm for the underlying quantum master
equation. We show how to evaluate this norm for different variational
classes of correlated and uncorrelated density matrices and exemplify
the application of our variational formalism using a dissipative Ising
model, which represents an important outstanding many-problem
connected to experiments with strongly interacting Rydberg
atoms. Importantly, our variational method is not limited to such
Ising models, which we demonstrate by providing a variational analysis
of the driven-dissipative Bose-Hubbard model. Finally, we conclude
with a comparison of our variational approach to a previously employed
method, establishing the advantages of our novel method.

The analysis of the stationary state is based on the quantum master
equation describing the dynamics of the density matrix $\rho$. In the
case of a Markovian evolution, it can be given in Lindblad form,
\begin{equation}
\frac{d}{dt}\rho = -i[H,\rho] + \sum\limits_{i} \left(c_i\rho c_i^\dagger - \frac{1}{2}\left\{c_i^\dagger c_i, \rho\right\}\right),
\label{eq:master}
\end{equation}
where $H$ is the Hamiltonian describing the coherent evolution and the
jump operators $c_i$ are associated with the incoherent part of the
dynamics \cite{Breuer2002}. The stationary states of the system can
then be obtained by solving the equation $d\rho/dt = 0$. Outside of
one-dimensional problems, where algorithms derived from the
density-matrix renormalization group can be applied
\cite{Honing2012,Honing2013,Pizorn2013}, the exact stationary state
cannot be computed for sufficiently large system sizes. In many
previous works
\cite{Diehl2010a,Tomadin2011,Liu2011,Lee2011,Lee2012a,Qian2013,Lee2013,Jin2013},
a mean-field approximation has been applied to study the stationary
state. To understand the key features of our variational method, it is
instructive to briefly restate the basic elements of this mean-field
approach. There, the partial trace over the whole system except a
single site is taken, resulting in the single site equations,
\begin{equation}
  \frac{d}{dt}\rho_i = \tr[\not{i}]{\frac{d}{dt}\rho} = -i[H_i,\rho_i] + \mathcal{D}_i(\rho_i).
  \label{eq:mf}
\end{equation}
where $H_i$ and $\mathcal{D}_i$ are the mean-field Hamiltonian and
dissipative terms, respectively \cite{Weimer2008b,Navez2010}. The
stationary state of these purely local master equations can be
expressed as a product of single site density matrices, $\rho =
\prod_i \rho_i$  and is found from the condition $d\rho_i/dt = 0$ by
solving the set of nonlinear equations.

Within our variational method, we take a different approach towards
finding an approximate solution to the stationary state. Here, we take
the full quantum master equation into account, i.e., without
performing any partial trace operation.
Instead, we approximate the true stationary state by a variational
density matrix. Naturally, the true stationary state will in general
be outside the space of possible density matrices that can be
generated by a certain choice of variational degrees of freedom. As
the crucial step, we approximate the true state by the variational
state that minimizes a suitable norm of the full quantum master
equation, i.e.,
\begin{equation}
  ||\dot{\rho}|| = ||-i[H,\rho] + \mathcal{D}(\rho)|| \to \min.
\end{equation}
At first glance, it might seem that the choice of the norm would be
rather arbitrary without an underlying physical functional such as the
(free) energy, but we will show that this is not the case. First, any
suitable norm has to be invariant under unitary transformations of the
form $\dot{\rho}\to U\dot{\rho}U^\dagger$, making it natural to focus
on vector norms for the set of eigenvalues of $\dot{\rho}$ (e.g.,
Schatten norms such as the trace norm, Frobenius norm, or spectral
norm). Second, the norm also has to be \emph{unbiased} in the sense it
does not favor certain classes of density matrices. For example, any
Schatten norm not satisfying linearity according to
$||\dot{\rho}||=||\lambda \dot{\rho}||/\lambda$ will, when applied to
the time derivative of product states, always favor the maximally
mixed state for sufficiently large system sizes \footnote{See
  Supplemental Material for a proof of the biasedness of Schatten
  norms violating the linearity condition, which includes
  Ref.~\cite{Bhatia1997}.}. These requirements single out the trace
norm,
\begin{equation}
  ||\dot{\rho}|| = \trtxt{|\dot{\rho}|}.
\end{equation}
In the following, we discuss how to evaluate this norm for different
classes of variational density matrices.

\emph{Product states.---} Similar to the mean-field approach, the
simplest variational state one can construct is a product state of
single site density matrices,
\begin{equation}
  \rho = \rho_p \equiv \mathcal{R} 1 = \prod\limits_i \rho_i
\end{equation}
where we have introduced the superoperator $\mathcal{R}$, which
transforms every occurrence of the identity $1_i$ into the density
matrix $\rho_i$. Then, the time derivative
according to the quantum master equation has the form
\begin{equation}
  \dot{\rho} = \sum\limits_i \mathcal{R}\dot{\rho}_i +\sum\limits_{\langle ij \rangle} \mathcal{R} \dot{C}_{ij},
  \label{eq:dotrho}
\end{equation}
where $C_{ij}$ is a matrix accounting for correlations between sites
$i$ and $j$ and the double sum runs over all terms that are connected
through interaction terms in the Hamiltonian or correlated jump
operators. Note that although the variational state is a product
state, the resulting dynamics will typically generate
correlations. Here, we also see that the previous mean-field approach
of solving the equation $\dot{\rho}_i=0$ only minimizes the first term
in the norm $||\dot{\rho}||$, while the second term is uncontrolled.

At this point, a numerically exact minimization of the norm
$||\dot{\rho}||$ is in general still a computationally intractable
problem. However, it is vital to stress that our variational
formulation represents the central step towards obtaining a good
approximation to the stationary state. Importantly, for a successful
application of a variational principle, we do not require an exact
solution, but only to obtain an \emph{upper bound} to the norm. Here,
we can obtain such an upper bound through the inequality
\begin{align}
  ||\dot{\rho}|| 
\leq \sum\limits_{\langle ij \rangle} \tr{|\mathcal{R} \left(\dot{\rho}_i\rho_j+\rho_i\dot{\rho}_j + \dot{C}_{ij}\right)|} = \sum\limits_{\langle ij\rangle} \trtxt{|\dot{\rho}_{ij}|},
\end{align}
where we have used the reduced two-site operators $\dot{\rho}_{ij} =
\tr[\not{i}\not{j}]{\dot{\rho}}$. Consequently, for a translationally
invariant system, it is sufficient to minimize the norm
$||\dot{\rho}_{ij}||$ of a single bond. Note that although only a
single bond has to be treated exactly, the interaction with the other
surrounding sites is still being accounted for on a mean-field level.

Interestingly, we point out that the solution to our variational
approach depends on the coordination number $z$, in contrast to
variational methods based on product states for equilibrium
systems. Furthermore, in the limit $z \to \infty$, we recover the
solution to the mean-field equation $\dot{\rho}_i = 0$, which then
also becomes the exact solution of the full quantum master equation.

\emph{Quantum and classical correlations.---} We can now extend the
class of variational states to go beyond product states. The first step in this direction is to include nearest-neighbor correlations of the form
\begin{equation}
  \rho = \rho_c \equiv \rho_p + \sum\limits_{\langle ij \rangle} \mathcal{R} C_{ij} + \sum\limits_{\langle ij\rangle \ne \langle kl\rangle}\mathcal{R}C_{ij}C_{kl}+\ldots,
\end{equation}
subject to the constraint that all two-site density matrices
$\rho_{ij}$ are positive, which guarantees the positivity of $\rho_c$
\cite{Navez2010}. Using the same procedure as in the case of product
states, we find
\begin{equation}
||\dot{\rho}|| \leq \sum\limits_{\langle ijk \rangle} || \dot{\rho}_{ijk}||,
\label{eq:var}
\end{equation}
i.e., the norm can be bounded from above by minimizing all
combinations involving three sites. Note that on triangular lattices,
it is necessary to distinguish three-site operators forming a loop
from their open counterparts.

More generally, the variational approach also allows for basically
arbitrary choices of density matrices, including long-range and
higher-order correlations as well as long-range interactions. While
the resulting variational functional will in general be more
complicated than Eq.~(\ref{eq:var}), the underlying variational
principle will be unchanged.

\emph{Dissipative Ising model.---} We now turn to the study of a
concrete model using our variational approach. We focus on a
dissipative extension of the Ising model, in which one of the spin
states is incoherently pumped into the other. This model has received
considerable interest recently because of its expected relevance for
ongoing experiments with ultracold Rydberg atoms
\cite{Carr2013,Malossi2014} and because of controversies surrounding
the phase diagram of the nonequilibrium steady state
\cite{Lee2011,Honing2013,Hoening2014,Marcuzzi2014}. The quantum
master equation of this model is given in the Lindblad form of
Eq.~(\ref{eq:master}), with the Hamiltonian following from an Ising
model with interaction strength $V$, in both a transverse field $g$
and a longitudinal field $h$,
\begin{equation}
  H = \frac{g}{2} \sum\limits_i \sigma_x^{(i)} + \frac{h}{2} \sum\limits_i \sigma_z^{(i)} + \frac{V}{4} \sum\limits_{\langle ij\rangle}  \sigma_z^{(i)}\sigma_z^{(j)}.
\end{equation}
The jump operators describing spin flips can be expressed as $c_i =
\sqrt{\gamma}\sigma_-$, with $\gamma$ being the decay rate of the up
spins. In the context of experiments with Rydberg atoms, the down spin
state corresponds to the atomic ground state, while the up spin state
refers to a Rydberg state. The interaction strength $V=C_6/a^6$
follows from a repulsive van der Waals interaction described by a
$C_6$ coefficient at the lattice spacing $a$ and transitions between
the two states are driven by laser fields with a Rabi frequency
$\Omega=g$ and a detuning $\Delta$ that is related to the longitudinal
field as $h = \Delta + zV/2$ \cite{Low2012}. Finally, $\gamma$ refers
to the radiative decay rate of the Rydberg state, which can be tuned
by laser coupling to other excited non-Rydberg states \cite{Zhao2012}.

\begin{figure}[tb]

\includegraphics{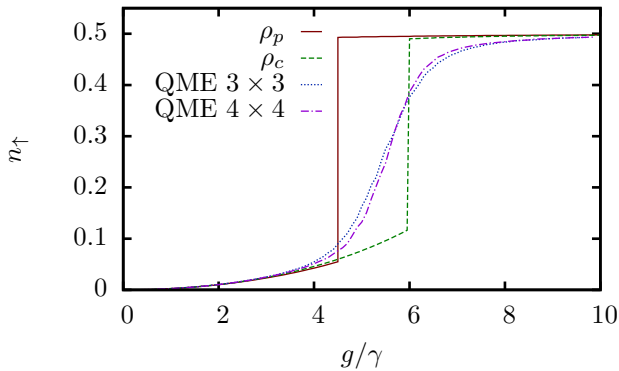}

\caption{Up-spin density $n_\uparrow$ in the steady state of the
  dissipative Ising model on a two-dimensional square lattice
  depending on the transverse-field $g$. Results are shown for the
  variational methods using product states $\rho_p$ and correlated
  states $\rho_c$, as well as solutions to the full quantum master
  equation (QME) for $3\times 3$ and $4 \times 4$ lattices ($h = 0$,
  $V=5\,\gamma$).}

\label{fig:corr2d}

\end{figure}

For this model, the mean-field solution of Eq.~(\ref{eq:mf}) predict a region of
bistability involving two distinct steady states over a large range of
the transverse field $g$ \cite{Lee2011,Hu2013,Marcuzzi2014}. This
finding is contradicted by numerical simulations for one-dimensional
systems based on the full quantum master equation
\cite{Ates2012a,Hu2013}. Here, our variational approach agrees with
the numerical results and always produces a unique steady state, both
for product states $\rho_p$ and for correlated states $\rho_c$. On a
two-dimensional square lattice, we find a first-order jump in the
up-spin density $n_\uparrow$, see Fig.~\ref{fig:corr2d}. Going from
the variational product state $\rho_p$ to the correlated state
$\rho_c$ results in a substantial increase of the critical value of
the first-order transition, from $g_c = 4.5$ to $g_c = 6.0$.

We complement our variational approach with numerical simulations
using a quantum trajectories method \cite{Daley2014,Johansson2013} for
$3\times 3$ and $4 \times 4$ lattices with periodic boundary
conditions. Due to finite-size effects, the first-order jump changes
into a smooth crossover, and the position of the peak the
susceptibility $\chi = \partial n_\uparrow/\partial g$ is shifted from
its infinite-system value \cite{Fisher1982}. Here, we find the
susceptibility peak at $g_{3\times 3} = 5.5$ and $g_{4\times 4} =
5.6$, respectively, implying good quantitative agreement between the
numerical results and the variational procedure for correlated states,
which is carried out in the thermodynamic limit of infinite system
size.

\begin{figure}[tb]

\includegraphics{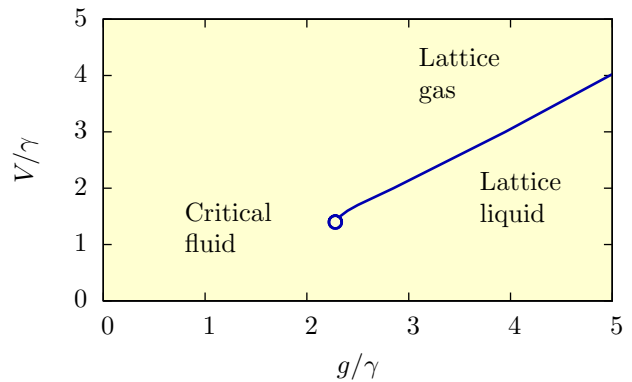}

\caption{Phase diagram of the dissipative transverse-field Ising model
  ($h=0$). For large interaction strength $V$, there is a first-order
  transition between a lattice gas and a lattice liquid of up
  spins. This first-order transition ends in a critical point, around
  which the system forms a critical fluid.}

\label{fig:phase}

\end{figure}

\emph{Phase diagram.---} We are now in the position to evaluate the
phase diagram of the dissipative Ising model, see
Fig.~\ref{fig:phase}. Here, the quantum master equation does not
exhibit any spin symmetry, as the dissipative terms break the
$\mathbb{Z}_2$ symmetry of the Ising model even for
$h=0$. Nevertheless, it is still possible to obtain phase transitions
if an emergent symmetry arises, as in the well known case of the
liquid-gas transition \cite{Huang1987}. Here, it is natural to treat
the up spins as particles, as then the dissipation can be understood
as a particle loss process. Then, we have a similar transition from a
low-density gas of up spins to a high-density liquid close to half
filling with almost vanishing compressibility $\kappa = -\partial
n_\uparrow/\partial h$ when the transverse field $g$ is increased over
a critical value.

For weakly interacting particles, it is possible to apply perturbation
theory in the interaction strength $V$, meaning that all observables
are analytical and there is no phase transition. Therefore, it is
naturally to associate the meaning of the inverse temperature $\beta$
in the liquid-gas transition to the interaction term $V$ in the
dissipative Ising model. Similar identifications can also be made on
the basis of the mean-field approach \cite{Marcuzzi2014}. As it is the
case in the classical liquid-gas transition, the variational approach
based on correlated states predicts that the transition between the
liquid and gas phases ends in a critical point, see
Fig.~\ref{fig:phase}, which we find here to be located at $(g, V) =
(2.28,1.40)\,\gamma$.

Within our variational approach, we always obtain a unique solution
for the stationary state and do not observe any regions of
bistability. Close to the first-order transition, however, we observe
two distinct states that both have a low residual dissipation and thus
correspond to the bistable states obtained from the solution of the
mean-field equations. Moreover, we would like to point out that the
bistability observed in one experiment \cite{Carr2013} could be
explained by a hysteresis behavior close to first-order phase
transitions \cite{Argawal1981} (see also discussion in
\cite{Marcuzzi2014}).

\emph{Driven-dissipative Bose-Hubbard model.---} To underline the
extensive flexibility of our approach, let us turn to a
driven-dissipative extension of the Bose-Hubbard model, which is
equally paradigmatic as the quantum Ising model in the context of
quantum many-body systems. Using bosonic creation and annihilation
operators $b_i^\dagger$ and $b_i$, respectively, its Hamiltonian is of
the form
\begin{equation}
  H = -J\sum\limits_{\langle i,j\rangle} b_i^\dagger b_j + \sum\limits_i \left[\frac{U}{2}n_i^2 - \mu n_i + F \left(b_i+b_i^\dagger\right)\right],
\end{equation}
where $J$ describes the inter-site hopping of bosons, $U$ accounts for
an on-site interaction involving the square of the density operator
$n_i = b^\dagger_ib_i$, $\mu$ is the chemical potential of the
particles, and $F$ describes a coherent driving term. The dissipative
term follows from a single-particle loss of the bosons, which is
described by jump operators of the form $c_i = \sqrt{\gamma} b_i$. The
driven-dissipative Bose-Hubbard model has been discussed in a wide
range of physical contexts, ranging from cavity and circuit QED arrays
to exciton-polaritons in semiconductor microcavities
\cite{Tomadin2010,Liew2013,LeBoite2013}.

\begin{figure}[tb]

\includegraphics[width=\linewidth]{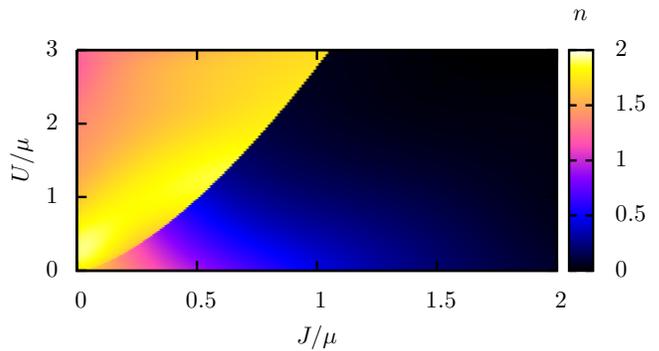}

\caption{Steady state density $n$ of the driven-dissipative
  Bose-Hubbard model ($F = 0.4\,\mu$, $\gamma = 0.2\,\mu$). The system
  undergoes a first-order jump in the density when the hopping $J$ is
  increased over a critical value. }

\label{fig:hubbard}

\end{figure}
Within our variational approach, we find a steady state solution for
the boson density as shown in Fig.~\ref{fig:hubbard}. We note that the
variational phase diagram differs from the mean-field prediction
\cite{LeBoite2013} in several points: (i) The steady state is always
unique, i.e., there is no bistability. (ii) There is a single
first-order transition between a high-density phase at small values of
the hopping $J$ and a low-density phase at large $J$. (iii) There are
no features reminiscent of Mott lobes in the steady state.

\emph{Comparison with the mean-field solution.---} Finally, we
conclude by making a detailed comparison of our variational principle
to the previously employed method of setting the mean-field equation
of motion Eq.~(\ref{eq:mf}) to zero. To ensure an unbiased comparison
on an equal footing, we carry out the variational procedure based on
product states $\rho_p$, as these states contain exactly the same free
parameters as in the mean-field method. To be explicit, we focus on
the case of the dissipative Ising model although the analysis can also
be applied interchangeably to the Bose-Hubbard model.

As noted earlier, the mean-field solution predicts extended regions of
bistable behavior, which is at odds both with numerical results and
our variational procedure. But even outside the bistable region, there
can be large qualitative and quantitative differences. In particular,
we find these for large negative values of the longitudinal field $h$,
where the mean-field solution predicts the existence of a large
antiferromagnetic phase \cite{Lee2011,Hu2013}. Within our variational
method, we find a drastic reduction of the extension of this phase.
This finding is also in line with numerical simulations on a $4 \times
4$ lattice, which suggests at least a strong suppression (if not
complete absence) of antiferromagnetic order.

\begin{figure}[t]

\includegraphics[width=\linewidth]{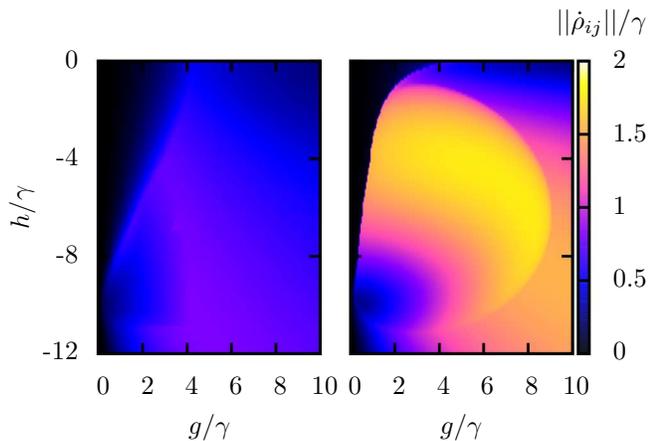}

\caption{Comparison of the norm of the two-site dynamics
  $||\dot{\rho}_{ij}||$ of the variational solution (left) and the
  mean-field solution (right) for the two-dimensional dissipative
  Ising model ($V=5\,\gamma$).}

\label{fig:comp}

\end{figure}

More quantitatively, we can express the difference between the two
methods by analyzing the norm $||\dot{\rho}_{ij}||$, expressing how
well the respective solutions approach the true stationary state of
the two-site dynamics. As expected, the variational method minimizing
this norm results in it being small throughout the $g-h$ plane, see
Fig.~\ref{fig:comp}. On the other hand, we find much larger values for
$||\dot{\rho}_{ij}||$ for the mean-field solution. In regions with
$||\dot{\rho}_{ij}|| > \gamma$, the contributions from the
correlations $\dot{C}_{ij}$ in Eq.~(\ref{eq:dotrho}) are comparable to
the spectral gap of the Liouvillian. This means the mean-field
solution is even inconsistent as these terms generating correlations
are of crucial importance and cannot be neglected as being only small
perturbations.

In summary, we have established and successfully applied a novel
variational principle for the non-equilibrium steady states of
driven-dissipative quantum many-body systems. Besides the systems
already studied in this work, our method will be of importance to all
settings in which interacting quantum many-body systems are coupled to
dissipative channels, in particular to exciton-polariton condensates
\cite{Kasprzak2006,Amo2009}, ultracold quantum gases in optical
cavities \cite{Baumann2010}, and Rydberg polaritons
\cite{Peyronel2012}.

\begin{acknowledgments}

  We acknowledge fruitful discussions with H.~P.~B\"uch\-ler,
  S.~Gopalakrishnan, N. Lang, T.~Lee, M. Lemeshko, I. Lesanovsky,
  E. Levi, M. Marcuzzi, and A.~Rapp. This work was funded by the
  Volkswagen Foundation.

\end{acknowledgments}



\begin{thebibliography}{10}

\bibitem{Kohn1999}
W.~Kohn,
\newblock Rev. Mod. Phys. {\bf 71}, 1253 (1999).

\bibitem{Schollwock2011}
U.~Schollw\"ock,
\newblock Ann. Phys. {\bf 326}, 96 (2011).

\bibitem{Diehl2008}
S.~Diehl, {A. Micheli}, {A. Kantian}, {B. Kraus}, {H. P. B\"uchler}, and {P.
  Zoller},
\newblock Nature Phys. {\bf 4}, 878 (2008).

\bibitem{Verstraete2009}
F.~Verstraete, M.~M. Wolf, and J.~Ignacio~Cirac,
\newblock Nature Phys. {\bf 5}, 633 (2009).

\bibitem{Weimer2010}
H.~{Weimer}, M.~{M{\"u}ller}, I.~{Lesanovsky}, P.~{Zoller}, and H.~P.
  {B{\"u}chler},
\newblock Nature Phys. {\bf 6}, 382 (2010).

\bibitem{Diehl2010}
S.~Diehl, W.~Yi, A.~J. Daley, and P.~Zoller,
\newblock Phys. Rev. Lett. {\bf 105}, 227001 (2010).

\bibitem{Alharbi2010}
A.~F. Alharbi and Z.~Ficek,
\newblock Phys. Rev. A {\bf 82}, 054103 (2010).

\bibitem{Barreiro2011}
J.~T. Barreiro, M.~M\"uller, P.~Schindler, D.~Nigg, T.~Monz, M.~Chwalla,
  M.~Hennrich, C.~F. Roos, P.~Zoller, and R.~Blatt,
\newblock Nature {\bf 470}, 486 (2011).

\bibitem{Krauter2011}
H.~Krauter, C.~A. Muschik, K.~Jensen, W.~Wasilewski, J.~M. Petersen, J.~I.
  Cirac, and E.~S. Polzik,
\newblock Phys. Rev. Lett. {\bf 107}, 080503 (2011).

\bibitem{Watanabe2012}
G.~Watanabe and H.~M\"akel\"a,
\newblock Phys. Rev. A {\bf 85}, 023604 (2012).

\bibitem{Rao2013}
D.~D.~B. Rao and K.~M\o{}lmer,
\newblock Phys. Rev. Lett. {\bf 111}, 033606 (2013).

\bibitem{Carr2013a}
A.~W. Carr and M.~Saffman,
\newblock Phys. Rev. Lett. {\bf 111}, 033607 (2013).

\bibitem{Weimer2013a}
H.~Weimer,
\newblock Mol. Phys. {\bf 111}, 1753 (2013).

\bibitem{Otterbach2014}
J.~Otterbach and M.~Lemeshko,
\newblock Phys. Rev. Lett. {\bf 113}, 070401 (2014).

\bibitem{Huber2012}
S.~D. Huber and H.~P. B\"uchler,
\newblock Phys. Rev. Lett. {\bf 108}, 193006 (2012).

\bibitem{Zhao2012}
B.~Zhao, A.~W. Glaetzle, G.~Pupillo, and P.~Zoller,
\newblock Phys. Rev. Lett. {\bf 108}, 193007 (2012).

\bibitem{Ates2012}
C.~Ates, B.~Olmos, W.~Li, and I.~Lesanovsky,
\newblock Phys. Rev. Lett. {\bf 109}, 233003 (2012).

\bibitem{Lemeshko2013a}
M.~{Lemeshko} and H.~{Weimer},
\newblock Nature Commun. {\bf 4}, 2230 (2013).

\bibitem{Kasprzak2006}
J.~Kasprzak, M.~Richard, S.~Kundermann, A.~Baas, P.~Jeambrun, J.~M.~J. Keeling,
  F.~M. Marchetti, M.~H. Szymańska, R.~André, J.~L. Staehli, V.~Savona, P.~B.
  Littlewood, B.~Deveaud, and L.~S. Dang,
\newblock Nature {\bf 443}, 409 (2006).

\bibitem{Amo2009}
A.~Amo, D.~Sanvitto, F.~P. Laussy, D.~Ballarini, E.~d. Valle, M.~D. Martin,
  A.~Lemaître, J.~Bloch, D.~N. Krizhanovskii, M.~S. Skolnick, C.~Tejedor, and
  L.~Viña,
\newblock Nature {\bf 457}, 291 (2009).

\bibitem{Hartmann2010}
M.~J. Hartmann,
\newblock Phys. Rev. Lett. {\bf 104}, 113601 (2010).

\bibitem{Baumann2010}
K.~{Baumann}, C.~{Guerlin}, F.~{Brennecke}, and T.~{Esslinger},
\newblock Nature {\bf 464}, 1301 (2010).

\bibitem{Nagy2010}
D.~Nagy, G.~K\'onya, G.~Szirmai, and P.~Domokos,
\newblock Phys. Rev. Lett. {\bf 104}, 130401 (2010).

\bibitem{Diehl2010a}
S.~Diehl, A.~Tomadin, A.~Micheli, R.~Fazio, and P.~Zoller,
\newblock Phys. Rev. Lett. {\bf 105}, 015702 (2010).

\bibitem{Tomadin2011}
A.~Tomadin, S.~Diehl, and P.~Zoller,
\newblock Phys. Rev. A {\bf 83}, 013611 (2011).

\bibitem{Lee2011}
T.~E. Lee, H.~H\"affner, and M.~C. Cross,
\newblock Phys. Rev. A {\bf 84}, 031402 (2011).

\bibitem{Kessler2012}
E.~M. Kessler, G.~Giedke, A.~Imamoglu, S.~F. Yelin, M.~D. Lukin, and J.~I.
  Cirac,
\newblock Phys. Rev. A {\bf 86}, 012116 (2012).

\bibitem{Honing2012}
M.~H\"oning, M.~Moos, and M.~Fleischhauer,
\newblock Phys. Rev. A {\bf 86}, 013606 (2012).

\bibitem{Honing2013}
M.~H\"oning, D.~Muth, D.~Petrosyan, and M.~Fleischhauer,
\newblock Phys. Rev. A {\bf 87}, 023401 (2013).

\bibitem{Horstmann2013}
B.~Horstmann, J.~I. Cirac, and G.~Giedke,
\newblock Phys. Rev. A {\bf 87}, 012108 (2013).

\bibitem{Torre2013}
E.~G.~D. Torre, S.~Diehl, M.~D. Lukin, S.~Sachdev, and P.~Strack,
\newblock Phys. Rev. A {\bf 87}, 023831 (2013).

\bibitem{Sieberer2013}
L.~M. Sieberer, S.~D. Huber, E.~Altman, and S.~Diehl,
\newblock Phys. Rev. Lett. {\bf 110}, 195301 (2013).

\bibitem{Qian2013}
J.~Qian, L.~Zhou, and W.~Zhang,
\newblock Phys. Rev. A {\bf 87}, 063421 (2013).

\bibitem{Lee2013}
T.~E. Lee, S.~Gopalakrishnan, and M.~D. Lukin,
\newblock Phys. Rev. Lett. {\bf 110}, 257204 (2013).

\bibitem{Carr2013}
C.~Carr, R.~Ritter, C.~G. Wade, C.~S. Adams, and K.~J. Weatherill,
\newblock Phys. Rev. Lett. {\bf 111}, 113901 (2013).

\bibitem{Joshi2013}
C.~Joshi, F.~Nissen, and J.~Keeling,
\newblock Phys. Rev. A {\bf 88}, 063835 (2013).

\bibitem{Malossi2014}
N.~Malossi, M.~M. Valado, S.~Scotto, P.~Huillery, P.~Pillet, D.~Ciampini,
  E.~Arimondo, and O.~Morsch,
\newblock Phys. Rev. Lett. {\bf 113}, 023006 (2014).

\bibitem{Marcuzzi2014}
M.~{Marcuzzi}, E.~{Levi}, S.~{Diehl}, J.~P. {Garrahan}, and I.~{Lesanovsky},
\newblock {a}rXiv:1406.1015  (2014).

\bibitem{Lang2014}
N.~{Lang} and H.~P. {B{\"u}chler},
\newblock {a}rXiv:1408.4616  (2014).

\bibitem{Breuer2002}
H.-P. Breuer and F.~Petruccione,
\newblock {\em The {T}heory of {O}pen {Q}uantum {S}ystems} (Oxford University
  Press, Oxford, 2002).

\bibitem{Pizorn2013}
I.~Pi\ifmmode~\check{z}\else \v{z}\fi{}orn,
\newblock Phys. Rev. A {\bf 88}, 043635 (2013).

\bibitem{Liu2011}
K.~Liu, L.~Tan, C.-H. Lv, and W.-M. Liu,
\newblock Phys. Rev. A {\bf 83}, 063840 (2011).

\bibitem{Lee2012a}
T.~E. Lee, H.~H\"affner, and M.~C. Cross,
\newblock Phys. Rev. Lett. {\bf 108}, 023602 (2012).

\bibitem{Jin2013}
J.~Jin, D.~Rossini, R.~Fazio, M.~Leib, and M.~J. Hartmann,
\newblock Phys. Rev. Lett. {\bf 110}, 163605 (2013).

\bibitem{Weimer2008b}
H.~Weimer, M.~J. Henrich, F.~Rempp, H.~Schr\"oder, and G.~Mahler,
\newblock Europhys. Lett. {\bf 83}, 30008 (2008).

\bibitem{Navez2010}
P.~Navez and R.~Sch\"utzhold,
\newblock Phys. Rev. A {\bf 82}, 063603 (2010).

\bibitem{Note1}
See Supplemental Material for a proof of the biasedness of Schatten norms
  violating the linearity condition, which includes Ref.~\cite{Bhatia1997}.

\makeatletter
\providecommand \@ifxundefined [1]{%
 \@ifx{#1\undefined}
}%
\providecommand \@ifnum [1]{%
 \ifnum #1\expandafter \@firstoftwo
 \else \expandafter \@secondoftwo
 \fi
}%
\providecommand \@ifx [1]{%
 \ifx #1\expandafter \@firstoftwo
 \else \expandafter \@secondoftwo
 \fi
}%
\providecommand \natexlab [1]{#1}%
\providecommand \enquote  [1]{``#1''}%
\providecommand \bibnamefont  [1]{#1}%
\providecommand \bibfnamefont [1]{#1}%
\providecommand \citenamefont [1]{#1}%
\providecommand \href@noop [0]{\@secondoftwo}%
\providecommand \href [0]{\begingroup \@sanitize@url \@href}%
\providecommand \@href[1]{\@@startlink{#1}\@@href}%
\providecommand \@@href[1]{\endgroup#1\@@endlink}%
\providecommand \@sanitize@url [0]{\catcode `\\12\catcode `\$12\catcode
  `\&12\catcode `\#12\catcode `\^12\catcode `\_12\catcode `\%12\relax}%
\providecommand \@@startlink[1]{}%
\providecommand \@@endlink[0]{}%
\providecommand \url  [0]{\begingroup\@sanitize@url \@url }%
\providecommand \@url [1]{\endgroup\@href {#1}{\urlprefix }}%
\providecommand \urlprefix  [0]{URL }%
\providecommand \Eprint [0]{\href }%
\providecommand \doibase [0]{http://dx.doi.org/}%
\providecommand \selectlanguage [0]{\@gobble}%
\providecommand \bibinfo  [0]{\@secondoftwo}%
\providecommand \bibfield  [0]{\@secondoftwo}%
\providecommand \translation [1]{[#1]}%
\providecommand \BibitemOpen [0]{}%
\providecommand \bibitemStop [0]{}%
\providecommand \bibitemNoStop [0]{.\EOS\space}%
\providecommand \EOS [0]{\spacefactor3000\relax}%
\providecommand \BibitemShut  [1]{\csname bibitem#1\endcsname}%
\let\auto@bib@innerbib\@empty
\bibitem [{\citenamefont {Bhatia}(1997)}]{Bhatia1997}%
  \BibitemOpen
  \bibfield  {author} {\bibinfo {author} {\bibfnamefont {R.}~\bibnamefont
  {Bhatia}},\ }\href {\doibase 10.1007/978-1-4612-0653-8} {\emph {\bibinfo
  {title} {Matrix analysis}}},\ \bibinfo {series} {Graduate Texts in
  Mathematics}, Vol.\ \bibinfo {volume} {169}\ (\bibinfo  {publisher}
  {Springer},\ \bibinfo {address} {New York},\ \bibinfo {year}
  {1997})\BibitemShut {NoStop}%
\bibitem{Hoening2014}
M.~Hoening, W.~Abdussalam, M.~Fleischhauer, and T.~Pohl,
\newblock Phys. Rev. A {\bf 90}, 021603 (2014).

\bibitem{Low2012}
R.~L\"ow, H.~Weimer, J.~Nipper, J.~B. Balewski, B.~Butscher, H.~P. B\"uchler,
  and T.~Pfau,
\newblock J. Phys. B {\bf 45}, 113001 (2012).

\bibitem{Hu2013}
A.~Hu, T.~E. Lee, and C.~W. Clark,
\newblock Phys. Rev. A {\bf 88}, 053627 (2013).

\bibitem{Ates2012a}
C.~Ates, B.~Olmos, J.~P. Garrahan, and I.~Lesanovsky,
\newblock Phys. Rev. A {\bf 85}, 043620 (2012).

\bibitem{Daley2014}
A.~J. Daley,
\newblock Adv. Phys. {\bf 63}, 77 (2014).

\bibitem{Johansson2013}
J.~Johansson, P.~Nation, and F.~Nori,
\newblock Comp. Phys. Comm. {\bf 184}, 1234 (2013).

\bibitem{Fisher1982}
M.~E. Fisher and A.~N. Berker,
\newblock Phys. Rev. B {\bf 26}, 2507 (1982).

\bibitem{Huang1987}
K.~Huang,
\newblock {\em Statistical Mechanics} (John Wiley and Sons, New York, 1987).

\bibitem{Argawal1981}
G.~S. Agarwal and S.~R. Shenoy,
\newblock Phys. Rev. A {\bf 23}, 2719 (1981).

\bibitem{Tomadin2010}
A.~Tomadin, V.~Giovannetti, R.~Fazio, D.~Gerace, I.~Carusotto, H.~E. T\"ureci,
  and A.~Imamoglu,
\newblock Phys. Rev. A {\bf 81}, 061801 (2010).

\bibitem{Liew2013}
T.~C.~H. Liew and V.~Savona,
\newblock New Journal of Physics {\bf 15}, 025015 (2013).

\bibitem{LeBoite2013}
A.~Le~Boit\'e, G.~Orso, and C.~Ciuti,
\newblock Phys. Rev. Lett. {\bf 110}, 233601 (2013).

\bibitem{Peyronel2012}
T.~{Peyronel}, O.~{Firstenberg}, Q.-Y. {Liang}, S.~{Hofferberth}, A.~V.
  {Gorshkov}, T.~{Pohl}, M.~D. {Lukin}, and V.~{Vuleti{\'c}},
\newblock Nature {\bf 488}, 57 (2012).

\end{thebibliography}

\end{document}




%
%

\title{Supplemental Material for ``Variational principle for steady
  states of dissipative quantum many-body systems''}

\author{Hendrik Weimer}%
\email{hweimer@itp.uni-hannover.de}
\affiliation{Institut f\"ur Theoretische Physik, Leibniz Universit\"at Hannover, Appelstr. 2, 30167 Hannover, Germany}



\maketitle

\section{Biasedness of Schatten norms}

We consider Schatten norms of the form
\begin{equation}
  ||\dot{\rho}|| = \trtxt{|\dot{\rho}|^p}.
\end{equation}
with $p \geq 1$. Note that we have defined the norm without taking the
$p$th root of the final result as this does not change the variational
minimum. Schatten norms form the most common class of norms invariant
under unitary transformations of the form $\dot{\rho}\to
U\dot{\rho}U^\dagger$ and corresponds to vector norms of the set of
eigenvalues of $\dot{\rho}$. Important examples include the trace norm
($p=1$), the Hilbert-Schmidt or Frobenius norm ($p=2$) and the spectral
norm ($p=\infty$). For $p<1$, the expression does not produce a valid
norm as the triangle inequality is violated \cite{Bhatia1997}. Here,
we consider the consequences of choosing a variational norm that
violates the linearity condition, i.e.,
\begin{equation}
  ||\dot{\rho}||\ne||\lambda \dot{\rho}||/\lambda,
\end{equation}
which is the case for all Schatten norms with $p>1$.

For product states of the form $\rho = \prod_i \rho_i$, we can write
the resulting variational norm as [cf.~Eq.~(6) of the main text]
\begin{equation}
  ||\dot{\rho}|| = ||\sum\limits_i \mathcal{R}\dot{\rho}_i +\sum\limits_{\langle ij \rangle} \mathcal{R} \dot{C}_{ij}|| \equiv ||\sum\limits_{\langle ij \rangle} \mathcal{R} A_{ij}||.
  \label{eq:dotrhosupp}
\end{equation}

Now, we are interested in obtaining an \emph{upper} bound for the norm,
which we will then evaluate for the maximally mixed state. We can use
the triangle inequality, resulting in
\begin{equation}
  ||\dot{\rho}|| \leq \sum\limits_{\langle ij\rangle} || A_{ij}|| \prod\limits_{k\ne ij} ||\rho_k||.
\end{equation}
As the sum contains on the order of $N$ terms, we can find a constant
$a$ such that
\begin{equation}
  ||\dot{\rho}|| \leq aN \prod\limits_{k>2} ||\rho_k||.
\end{equation}
For the maximally mixed state, this yields
\begin{equation}
  ||\dot{\rho}|| \leq \frac{aN}{d^{(p-1)(N-2)}},
  \label{eq:mms}
\end{equation}
where $d$ is the Hilbert space dimension of a single site. Here, we
already see that the trace norm with $p=1$ is special, as it is the
only Schatten norm not decaying exponentially with the system size $N$.

Next, we calculate a \emph{lower} bound for the norm of
Eq.~(\ref{eq:dotrhosupp}). Here, it is important to note that the
matrices $A_{ij}$ cannot be zero as the true stationary state cannot
be described by a product state for finite interactions and finite
spatial dimensions \cite{Navez2010}. As the $A_{ij}$ act on different
parts of the Hilbert space, their sum cannot vanish either and we can
find another constant such that
\begin{equation}
  ||\dot{\rho}|| \geq b \prod\limits_{k>2} ||\rho_k||.
\end{equation}
If we choose a variational state different from the maximally mixed
state, the largest eigenvalue of $\rho_k$, $p_{max}$, will satisfy the
relation $p_{max} > 1/d$. Without loss of generality, we assume
$p_{max}$ to be the same for all sites, which yields
\begin{equation}
||\dot{\rho}|| \geq b \left[p_{max}^p + \frac{d-2}{d^p} + \left(\frac{2}{d}-p_{max}\right)^p\right]^{N-2}.
\end{equation}
This expression again describes an exponential decay in $N$ for any
$p>1$, but with a strictly smaller base than for the upper bound of
Eq.~(\ref{eq:mms}) for the maximally mixed state. Consequently, there
will be a finite value of $N$ above which the maximally mixed state
will always produce a lower norm than any other state, proving
the biasedness of Schatten norms with $p>1$.

%